# Cluster Decomposition Principle and Two-Electron Wave Function of the Cooper Pair in the BCS Superconducting State

Katsuhiko Higuchi[1] and Masahiko Higuchi[2]

[1] *Graduate School of Advanced Sciences of Matter, Hiroshima University, Higashi-Hiroshima 739-8527, Japan*

[2] *Department of Physics, Faculty of Science, Shinshu University, Matsumoto 390-8621, Japan*

**Abstract**

We present the explicit forms of the maximum eigenvalue and the corresponding eigenfunction for the second-order reduced density matrix (RDM2) of the BCS superconducting state (SS). Using these quantities, we deal with two topics in the present paper. As the first topic, it is shown that the cluster decomposition principle holds in the BCS-SS. This proof gives a theoretical foundation that the abnormal density can be chosen as the order parameter of the SS. As the second topic, it is shown that such an eigenfunction is spin singlet and spatially extends isotopically, and further that the mean distance of two electrons which consists of the above eigenfunction is in a good agreement with Pippard's coherence length. This means that maximum geminal of the RDM2 of the BCS-SS can be regarded as the Cooper pair itself which are condensed to the same energy level in a number of $O(N)$.

Key words:

superconductivity, superconducting state, second-order reduced density matrix, geminal, cluster decomposition principle, fluctuation of the particle number, order parameter, Pippard's coherence length

## 1. Introduction

In a lot of first-principles calculations of the superconductivity, the abnormal density $\langle \Phi | \psi(\mathbf{r}'\zeta')\psi(\mathbf{r}\zeta) | \Phi \rangle$ is generally regarded as an order parameter of the superconducting state (SS), where $\psi(\mathbf{r}\zeta)$ is the field operator of electrons and $|\Phi\rangle$ denotes a SS [1, 2, 3, 4]. In particular, the density functional theory for the superconductor (SC-DFT) works well when the abnormal density is chosen as one of basic variables [5-36]. To be more specific, the temperature where the abnormal density disappears can be regarded as a critical temperature of the SS, and such a critical temperature has been estimated quantitatively well by the SC-DFT [5-36].

The definition of the superconductivity is given on the basis of the Bose-Einstein condensation (BEC) of the fermion system [37,38]. According to this definition, when the system in the SS, the magnitude of the maximum eigenvalue of the second-order reduced density matrix (RDM2) is $O(N)$ and the maximum geminal belongs to type (b) defined in the previous paper [39]. Here note that the maximum geminal of the RDM2 is defined as the eigenfunction of the RDM2 with the maximum eigenvalue. In the previous paper, it has been pointed out that the abnormal density can be an OP of the SS only if the cluster decomposition principle holds [39]. However, it is not obvious whether the cluster decomposition principle holds or not in the SS. In other words, the cluster decomposition principle has not well argued in the SSs including the BCS-SS, so far. Therefore, the suspicion whether the abnormal density is appropriate as the order parameter of the SS has remained persistent up to the present even at the level of BCS-SS. In this paper, we clear this persistent suspicion at the level of BCS-SS. That is to say, the cluster decomposition principle is proven true in the BCS-SS. Using this proof, we can say that the abnormal density corresponds to the maximum geminal of the RDM2 of the BCS-SS multiplied by the square root of twice the number of maximum geminal in the SS.

In the previous paper [39], it has been also shown that the maximum geminal is condensed to one level in a number of $O(N)$ when the system is in the SS. In this paper, it is shown that the maximum geminal is spin singlet and has s-wave symmetry, and further that the mean distance of electrons which consist of the maximum geminal is in a good agreement with Pippard's coherence length of the BCS theory [40-43]. This means that the maximum geminal of the RDM2 of the BCS-SS corresponds to the Cooper pair itself of the BCS-SS.

Thus, we give answers to the fundamental questions " Is the abnormal density appropriate for the OP of the SS?", and "what is the relation between the maximum geminal and Cooper pair?" in the case of the BCS-SS, which is the most typical case of the SS's.

## 2. Cluster decomposition principle in the BCS superconducting state

First, we shall prove that the cluster decomposition principle holds in the BCS-SS. The proof is performed in the following 5 steps.

(1) Notations for various physical quantities are given for the later convenience.
(2) The cluster decomposition principle is reviewed.
(3) The RDM2 of the BCS-SS is calculated.
(4) The eigenvalue and eigenfunction for the RDM2 of the BCS-SS are found.
(5) The expectation value of the abnormal density with respect to the BCS-SS is calculated. The result of Step (4) and that of Step (5) are compared to each other, so that it is confirmed that the cluster decomposition principle holds in the BCS-SS [39].

*Step (1)*

We shall define the notations of various physical quantities as a preparation. A free electron state with the wavenumber $\boldsymbol{k}$ and spin $\sigma\ (=\uparrow,\downarrow)$ is denoted as $|\boldsymbol{k}\sigma\rangle$. Using the creation operator $C^{\dagger}_{\boldsymbol{k}\sigma}$, the state $|\boldsymbol{k}\sigma\rangle$ is written as

$$|\boldsymbol{k}\sigma\rangle = C^{\dagger}_{\boldsymbol{k}\sigma}|0\rangle\,, \tag{1}$$

where $|0\rangle$ is the vacuum state. Similarly, two electron state $|\boldsymbol{k}_1\sigma_1,\boldsymbol{k}_2\sigma_2\rangle$ is defined as

$$|\boldsymbol{k}_1\sigma_1,\boldsymbol{k}_2\sigma_2\rangle = C^{\dagger}_{\boldsymbol{k}_1\sigma_1}C^{\dagger}_{\boldsymbol{k}_2\sigma_2}|0\rangle\,. \tag{2}$$

As a special case of equation (2), two electron state $|\boldsymbol{k}\uparrow,-\boldsymbol{k}\downarrow\rangle$ is written as

$$|\boldsymbol{k}\uparrow,-\boldsymbol{k}\downarrow\rangle = B^{\dagger}_{\boldsymbol{k}}|0\rangle\,, \tag{3}$$

where the creation operator of two electron state $B^{\dagger}_{\boldsymbol{k}}$ is defined as $B^{\dagger}_{\boldsymbol{k}} = C^{\dagger}_{\boldsymbol{k}\uparrow}C^{\dagger}_{-\boldsymbol{k}\downarrow}$. The coordinate representation of equation (3) is denoted as $\langle \boldsymbol{r}\zeta,\boldsymbol{r}'\zeta'|\boldsymbol{k}\uparrow,-\boldsymbol{k}\downarrow\rangle$. Here the state $|\boldsymbol{r}\zeta,\boldsymbol{r}'\zeta'\rangle$ is also two electron state which is defined as

$$|\boldsymbol{r}\zeta,\boldsymbol{r}'\zeta'\rangle = \psi^{\dagger}(\boldsymbol{r}\zeta)\psi^{\dagger}(\boldsymbol{r}'\zeta')|0\rangle\,, \tag{4}$$

where $\psi(\boldsymbol{r}\zeta)$ and $\psi^{\dagger}(\boldsymbol{r}\zeta)$ are the field operators of electrons. The operator of the RDM2 is defined as [39, 44]

$$\hat{\rho}^{(2)}(\boldsymbol{r}_1\zeta_1\boldsymbol{r}_2\zeta_2,\boldsymbol{r}_1'\zeta_1'\boldsymbol{r}_2'\zeta_2') = \frac{1}{2}\psi^{\dagger}(\boldsymbol{r}_1'\zeta_1')\psi^{\dagger}(\boldsymbol{r}_2'\zeta_2')\psi(\boldsymbol{r}_2\zeta_2)\psi(\boldsymbol{r}_1\zeta_1)\,. \tag{5}$$

*Step (2)*

If there does not exist any correlation between two systems which separate from each other in an infinite distance, the cluster decomposition principle holds [39, 45]. Specifically, we consider applying the following limits to the RDM2 [39]:

$$\boldsymbol{r}_1 \approx \boldsymbol{r}_2,\ \boldsymbol{r}_1' \approx \boldsymbol{r}_2',\ \text{and}\ |\boldsymbol{r}_1 - \boldsymbol{r}_1'| \to \infty. \tag{6}$$

If the cluster decomposition principle holds in a state $|\Phi\rangle$, then the expectation value of the RDM2 is expressed as the product of expectation values of abnormal densities. Namely, we have [39]

$$\langle\Phi|\hat{\rho}^{(2)}(\boldsymbol{r}_1\zeta_1\boldsymbol{r}_2\zeta_2,\boldsymbol{r}_1'\zeta_1'\boldsymbol{r}_2'\zeta_2')|\Phi\rangle \underset{Eq.\,(6)}{\longrightarrow} \frac{1}{2}\langle\Phi|\psi^{\dagger}(\boldsymbol{r}_1'\zeta_1')\psi^{\dagger}(\boldsymbol{r}_2'\zeta_2')|\Phi\rangle\langle\Phi|\psi(\boldsymbol{r}_2\zeta_2)\psi(\boldsymbol{r}_1\zeta_1)|\Phi\rangle\,. \tag{7}$$

Equation (7) is called the cluster decomposition principle of the RDM2. As mentioned above, it is not well argued whether Eq. (7) holds or not when $|\Phi\rangle$ is the SS.

*Step (3)*

An explicit form of the BCS wave function $|\Theta_{\mathrm{BCS}}\rangle$ is given in Appendix A [42]. Let us consider the expectation value of equation (5) with respect to $|\Theta_{\mathrm{BCS}}\rangle$. It is calculated as

$$\begin{aligned}\rho^{(2)}_{\Theta_{\mathrm{BCS}}}(\boldsymbol{r}_1\zeta_1\boldsymbol{r}_2\zeta_2,\boldsymbol{r}_1'\zeta_1'\boldsymbol{r}_2'\zeta_2') = \frac{1}{2}\sum_{\boldsymbol{k}}\sum_{\boldsymbol{k}'}\Big\{&\langle\boldsymbol{r}_1'\zeta_1'\boldsymbol{r}_2'\zeta'|\boldsymbol{k}'\uparrow,-\boldsymbol{k}'\downarrow\rangle^{*}\langle\boldsymbol{r}_1\zeta_1\boldsymbol{r}_2\zeta_2|\boldsymbol{k}\uparrow,-\boldsymbol{k}\downarrow\rangle\\ &\times v_{\boldsymbol{k}'}^{*}v_{\boldsymbol{k}}\langle\tilde{\Theta}^{\boldsymbol{k}'}_{\mathrm{BCS}}|\tilde{\Theta}^{\boldsymbol{k}}_{\mathrm{BCS}}\rangle\Big\},\end{aligned} \tag{8}$$

where $v_{\boldsymbol{k}}$ represents the probability amplitude such that $|v_{\boldsymbol{k}}|^2$ gives the probability of two electron state $\left|\boldsymbol{k}\uparrow,-\boldsymbol{k}\downarrow\right\rangle$ being occupied in the BCS-SS, and where the state $\left|\tilde{\Theta}_{\mathrm{BCS}}^{\boldsymbol{k}}\right\rangle$ is the state in which $\left|\boldsymbol{k}\uparrow,-\boldsymbol{k}\downarrow\right\rangle$ is excluded from $\left|\Theta_{\mathrm{BCS}}\right\rangle$. The inner product of $\left|\tilde{\Theta}_{\mathrm{BCS}}^{\boldsymbol{k}}\right\rangle$ and $\left|\tilde{\Theta}_{\mathrm{BCS}}^{\boldsymbol{k}'}\right\rangle$, which appears in equation (8), is calculated as

$$\left\langle\tilde{\Theta}_{\mathrm{BCS}}^{\boldsymbol{k}'}\middle|\tilde{\Theta}_{\mathrm{BCS}}^{\boldsymbol{k}}\right\rangle = u_{\boldsymbol{k}}{}^{*}u_{\boldsymbol{k}'} + v_{\boldsymbol{k}}{}^{*}v_{\boldsymbol{k}'}\delta_{\boldsymbol{k}\boldsymbol{k}'}\,. \tag{9}$$

where $u_{\boldsymbol{k}}$ represents the probability amplitude such that $\left|u_{\boldsymbol{k}}\right|^2$ gives the probability of two electron state $\left|\boldsymbol{k}\uparrow,-\boldsymbol{k}\downarrow\right\rangle$ being unoccupied in the BCS-SS. Details of derivations of equations (8) and (9) are given in Appendix A.

*Step (4)*

We shall consider the eigenvalue and eigenfunction of equation (8). The eigenvalue equation of equation (8) is written as [39, 44]

$$\sum_{\zeta_1'}\sum_{\zeta_2'}\int d^3r_1'\int d^3r_2'\rho_{\Theta_{\mathrm{BCS}}}^{(2)}(\boldsymbol{r}_1\zeta_1\boldsymbol{r}_2\zeta_2,\boldsymbol{r}_1'\zeta_1'\boldsymbol{r}_2'\zeta_2')\left\langle\boldsymbol{r}_1'\zeta_1'\boldsymbol{r}_2'\zeta_2'\middle|\theta_{\Theta_{\mathrm{BCS}}}\right\rangle = n_\theta^{(2)}\left\langle\boldsymbol{r}_1\zeta_1\boldsymbol{r}_2\zeta_2\middle|\theta_{\Theta_{\mathrm{BCS}}}\right\rangle, \tag{10}$$

where $n_\theta^{(2)}$ and $\left\langle\boldsymbol{r}_1\zeta_1\boldsymbol{r}_2\zeta_2\middle|\theta_{\Theta_{\mathrm{BCS}}}\right\rangle$ are eigenvalue and eigenfunction for $\rho_{\Theta_{\mathrm{BCS}}}^{(2)}(\boldsymbol{r}_1\zeta_1\boldsymbol{r}_2\zeta_2,\boldsymbol{r}_1'\zeta_1'\boldsymbol{r}_2'\zeta_2')$, respectively. The maximum eigenvalue and the corresponding eigenfunction can be found by utilizing equation (3). The result obtained is

$$\left|\theta_{\Theta_{\mathrm{BCS}}}^{\max}\right\rangle = \sum_{\boldsymbol{k}} u_{\boldsymbol{k}}v_{\boldsymbol{k}}B_{\boldsymbol{k}}^{\dagger}\left|0\right\rangle, \tag{11}$$

where $\left|\theta_{\Theta_{\mathrm{BCS}}}^{\max}\right\rangle$ is the eigenfunction which yields the maximum eigenvalue, which is hereafter called the *maximum geminal* of the RDM2 of the BCS-SS. As shown below, it can be confirmed that equation (11) actually gives the maximum eigenvalue. Substituting equations (8), (9) and (11) into the LHS of equation (10), we get

$$\sum_{\zeta_1'}\sum_{\zeta_2'}\int d^3r_1'\int d^3r_2'\rho^{(2)}_{\Theta_{\rm BCS}}(\boldsymbol{r}_1\zeta_1\boldsymbol{r}_2\zeta_2,\boldsymbol{r}_1'\zeta_1'\boldsymbol{r}_2'\zeta_2')\left\langle \boldsymbol{r}_1'\zeta_1'\boldsymbol{r}_2'\zeta_2' \middle| \theta^{\max}_{\Theta_{\rm BCS}}\right\rangle$$
$$=\sum_{\boldsymbol{k}}\left\{\sum_{\boldsymbol{k}'}|v_{\boldsymbol{k}'}|^2 u_{\boldsymbol{k}'}{}^2+|v_{\boldsymbol{k}}|^4\right\}u_{\boldsymbol{k}}v_{\boldsymbol{k}}\left\langle \boldsymbol{r}_1\zeta_1\boldsymbol{r}_2\zeta_2 \middle| \boldsymbol{k}\uparrow,-\boldsymbol{k}\downarrow\right\rangle, \tag{12}$$

where equation (A-6) shown in Appendix A is used. In the RHS of equation (12), the magnitude of $|v_{\boldsymbol{k}}|^4$ is less than or equal to unity while that of $\sum_{\boldsymbol{k}'}|v_{\boldsymbol{k}'}|^2 u_{\boldsymbol{k}'}{}^2$ is shown to be $O(N)$. The proof is given in Appendix B. Therefore equation (12) is reasonably approximated as

$$\sum_{\zeta_1'}\sum_{\zeta_2'}\int d^3r_1'\int d^3r_2'\rho^{(2)}_{\Theta_{\rm BCS}}(\boldsymbol{r}_1\zeta_1\boldsymbol{r}_2\zeta_2,\boldsymbol{r}_1'\zeta_1'\boldsymbol{r}_2'\zeta_2')\left\langle \boldsymbol{r}_1'\zeta_1'\boldsymbol{r}_2'\zeta_2' \middle| \theta^{\max}_{\Theta_{\rm BCS}}\right\rangle \approx \left\{\sum_{\boldsymbol{k}'}|v_{\boldsymbol{k}'}|^2 u_{\boldsymbol{k}'}{}^2\right\}\left\langle \boldsymbol{r}_1\zeta_1\boldsymbol{r}_2\zeta_2 \middle| \theta^{\max}_{\Theta_{\rm BCS}}\right\rangle, \tag{13}$$

where equation (11) is used. Since the magnitude of the eigenvalue for the RDM2 means the occupation number of two-particle states [37, 38], and since the magnitude of the eigenvalue for equation (13) is $O(N)$ (see Appendix B), $\sum_{\boldsymbol{k}'}|v_{\boldsymbol{k}'}|^2 u_{\boldsymbol{k}'}{}^2$ corresponds to the maximum eigenvalue of the RDM2 of the BCS-SS. Namely, equation (13) is an eigenvalue equation of RDM2 of the BCS-SS with the maximum eigenvalue and the corresponding eigenstate.

Considering the normalization constant of the eigenfunction, the eigenvalue equation (13) is rewritten as

$$\sum_{\zeta_1'}\sum_{\zeta_2'}\int d^3r_1'\int d^3r_2'\rho^{(2)}_{\Theta_{\rm BCS}}(\boldsymbol{r}_1\zeta_1\boldsymbol{r}_2\zeta_2,\boldsymbol{r}_1'\zeta_1'\boldsymbol{r}_2'\zeta_2')\nu^{\max}_{\Theta_{\rm BCS}}(\boldsymbol{r}_1'\zeta_1',\boldsymbol{r}_2'\zeta_2')=n^{(2)}_{\max}\nu^{\max}_{\Theta_{\rm BCS}}(\boldsymbol{r}_1\zeta_1,\boldsymbol{r}_2\zeta_2), \tag{14}$$

where

$$n^{(2)}_{\max}=\sum_{\boldsymbol{k}'}|v_{\boldsymbol{k}'}|^2 u_{\boldsymbol{k}'}{}^2, \tag{15}$$

$$\nu^{\max}_{\Theta_{\rm BCS}}(\boldsymbol{r}_1\zeta_1,\boldsymbol{r}_2\zeta_2)=\frac{1}{\sqrt{2n^{(2)}_{\max}}}\left\langle \boldsymbol{r}_1\zeta_1\boldsymbol{r}_2\zeta_2 \middle| \theta^{\max}_{\Theta_{\rm BCS}}\right\rangle. \tag{16}$$

The BCS-SS can be regarded as the BEC of the fermions which consists of maximum geminals $\nu^{\max}_{\Theta_{\rm BCS}}(\boldsymbol{r}_1\zeta_1,\boldsymbol{r}_2\zeta_2)$ in a number of $O(N)$.

*Step (5)*

In this step, we shall show that the BCS-SS satisfies the cluster decomposition principle. Using equations (A-1) and (A-4) shown in Appendix A, the expectation value of the abnormal density with respect to the BCS-SS is given by

$$\begin{aligned}\langle\Theta_{\rm BCS}|\psi(\boldsymbol{r}'\zeta')\psi(\boldsymbol{r}\zeta)|\Theta_{\rm BCS}\rangle &= \left\{\prod_{\boldsymbol{k}'}\frac{1}{1+|\alpha_{\boldsymbol{k}'}|^2}\right\}\sum_{\boldsymbol{k}}\langle\boldsymbol{r}\zeta\boldsymbol{r}'\zeta'|\boldsymbol{k}\uparrow,-\boldsymbol{k}\downarrow\rangle\alpha_{\boldsymbol{k}} \\ &\quad\times\langle 0|\prod_{\boldsymbol{k}'''}\left(1+\alpha_{\boldsymbol{k}'''}{}^{*}B_{\boldsymbol{k}'''}\right)\prod_{\boldsymbol{k}''(\neq\boldsymbol{k})}\left(1+\alpha_{\boldsymbol{k}''}B_{\boldsymbol{k}''}{}^{\dagger}\right)|0\rangle.\end{aligned} \tag{17}$$

Using the commutation relations of the operators $B_{\boldsymbol{k}}$ and $B_{\boldsymbol{k}}{}^{\dagger}$, equation (17) is rigorously calculated to be

$$\langle\Theta_{\rm BCS}|\psi(\boldsymbol{r}'\zeta')\psi(\boldsymbol{r}\zeta)|\Theta_{\rm BCS}\rangle = \sum_{\boldsymbol{k}}u_{\boldsymbol{k}}v_{\boldsymbol{k}}\langle\boldsymbol{r}\zeta\boldsymbol{r}'\zeta'|\boldsymbol{k}\uparrow,-\boldsymbol{k}\downarrow\rangle. \tag{18}$$

Further using equations (11) and (15), the above equation finally becomes

$$\langle\Theta_{\rm BCS}|\psi(\boldsymbol{r}'\zeta')\psi(\boldsymbol{r}\zeta)|\Theta_{\rm BCS}\rangle = \sqrt{2n^{(2)}_{\max}}\,\nu^{\max}_{\Theta_{\rm BCS}}(\boldsymbol{r}\zeta,\boldsymbol{r}'\zeta'). \tag{19}$$

On the other hand, if the cluster decomposition principle holds for the state $|\Phi\rangle$, then equation (7) holds. Using this fact and further using the spectrum decomposition of the RDM2 [39], it has been shown in the previous paper that if the cluster decomposition principle holds for the state $|\Phi\rangle$, then the following relation is established [39]:

$$\langle\Phi|\psi(\boldsymbol{r}'\zeta')\psi(\boldsymbol{r}\zeta)|\Phi\rangle = \sqrt{2n^{(2)}_{\max}}\,\nu^{\max}_{\Phi}(\boldsymbol{r}\zeta,\boldsymbol{r}'\zeta'), \tag{20}$$

where $|\Phi\rangle$ is the general SS not limited to the BCS-SS, and where $\nu_{\Phi}^{\max}(\boldsymbol{r}\zeta,\boldsymbol{r}'\zeta')$ is the maximum geminal of the RDM2 of $|\Phi\rangle$ [39]. By employing this result, it is concluded from equation (19) that the cluster decomposition principle holds in the BCS-SS $|\Theta_{\mathrm{BCS}}\rangle$.

## 3. Features of the maximum geminal

We shall rewrite the maximum geminal $\nu_{\Theta_{\mathrm{BCS}}}^{\max}(\boldsymbol{r}'\zeta',\boldsymbol{r}\zeta)$ by using the relative coordinates, $\boldsymbol{\rho}=\boldsymbol{r}-\boldsymbol{r}'$, and center of mass coordinates, $\boldsymbol{R}=(\boldsymbol{r}+\boldsymbol{r}')/2$. Substituting equation (11) into equation (16), we get

$$\nu_{\Theta_{\mathrm{BCS}}}^{\max}(\boldsymbol{r}\zeta,\boldsymbol{r}'\zeta')=\frac{1}{\sqrt{2n_{\max}^{(2)}}}\frac{1}{\Omega}\sum_{\boldsymbol{k}}u_{\boldsymbol{k}}v_{\boldsymbol{k}}\left\{e^{i\boldsymbol{k}\cdot\boldsymbol{\rho}}\chi_{\uparrow}(\zeta)\chi_{\downarrow}(\zeta')-e^{-i\boldsymbol{k}\cdot\boldsymbol{\rho}}\chi_{\uparrow}(\zeta')\chi_{\downarrow}(\zeta)\right\}, \tag{21}$$

where $\Omega$ is the volume of the system. Note that $\nu_{\Theta_{\mathrm{BCS}}}^{\max}(\boldsymbol{r}'\zeta',\boldsymbol{r}\zeta)$ is not dependent on $\boldsymbol{R}$ but is dependent only on the relative coordinates $\boldsymbol{\rho}$, which means that the maximum geminal extend homogeneously in the system.

If we consider the case where the attractive interaction between electrons is isotropic, then the gap parameter is independent of the wavevector, i.e., the gap parameter is approximated as $\Delta_0$ (=constant) [42, 43]. In this case, it can be shown that the maximum geminal $\nu_{\Theta_{\mathrm{BCS}}}^{\max}(\boldsymbol{r}'\zeta',\boldsymbol{r}\zeta)$ is spin singlet and s-wave state. Actually, equation (21) is rewritten as

$$\nu_{\Theta_{\mathrm{BCS}}}^{\max}(\boldsymbol{r}\zeta,\boldsymbol{r}'\zeta')=\Xi_S(\zeta,\zeta')\frac{1}{4\pi^2|\boldsymbol{\rho}|\sqrt{n_{\max}^{(2)}}}\int dk\left(\frac{\Delta_0^{\,2}}{\xi_{\boldsymbol{k}}^{\,2}+\Delta_0^{\,2}}\right)^{\frac{1}{2}}k\sin(k|\boldsymbol{\rho}|), \tag{22}$$

where $\xi_{\boldsymbol{k}}$ is the free-electron energy shifted by the chemical potential $\mu$, i.e., $\xi_{\boldsymbol{k}}=\hbar^2k^2/(2m)-\mu$, and where $\Xi_S(\zeta,\zeta')$ is the spin singlet wave function given by

$$\Xi_S(\zeta,\zeta')=\frac{1}{\sqrt{2}}\left\{\chi_{\uparrow}(\zeta)\chi_{\downarrow}(\zeta')-\chi_{\uparrow}(\zeta')\chi_{\downarrow}(\zeta)\right\}. \tag{23}$$

The maximum geminal $v_{\Theta_{\rm BCS}}^{\max}\left(\boldsymbol{r}'\zeta',\boldsymbol{r}\zeta\right)$ is dependent on the magnitude of $\boldsymbol{\rho}$ but is not dependent on the angular coordinates of $\boldsymbol{\rho}$. This means that the maximum geminal $v_{\Theta_{\rm BCS}}^{\max}\left(\boldsymbol{r}'\zeta',\boldsymbol{r}\zeta\right)$ is exactly s-wave state. Details of the derivation of equation (22) is shown in Appendix C.

**4. Mean distance of two electrons in the maximum geminal**

The mean-square-distance between two electrons which consist of the maximum geminal $\left|v_{\Theta_{\rm BCS}}^{\max}\right\rangle$ is given by

$$\left\langle|\boldsymbol{\rho}|^2\right\rangle=\frac{\sum_{\zeta_1}\sum_{\zeta_2}\int v_{\Theta_{\rm BCS}}^{\max}\left(\boldsymbol{\rho},\zeta_1,\zeta_2\right)^*|\boldsymbol{\rho}|^2\,v_{\Theta_{\rm BCS}}^{\max}\left(\boldsymbol{\rho},\zeta_1,\zeta_2\right)d^3\rho}{\sum_{\zeta_1}\sum_{\zeta_2}\int\left|v_{\Theta_{\rm BCS}}^{\max}\left(\boldsymbol{\rho},\zeta_1,\zeta_2\right)\right|^2d^3\rho}. \tag{24}$$

Substituting equation (21) into equation (24), we get the following result after careful calculations:

$$\left\langle|\boldsymbol{\rho}|^2\right\rangle=\frac{\hbar^2{v_F}^2}{8{\Delta_0}^2}, \tag{25}$$

where $v_F$ is the Fermi velocity. In the derivation of equation (25), we consider the case where the attractive interaction between electrons is isotropic, and employ an approximation $\arctan\left(-\mu/\Delta_0\right)\approx-\pi/2$ because $\Delta_0\ll\mu$ holds. Details of derivation of equation (25) are given in Appendix D.

Therefore, the root-mean-square-distance (which is abbreviated as *mean distance*) between two electrons for the maximum geminal is given by

$$\bar{\rho}=\sqrt{\left\langle|\boldsymbol{\rho}|^2\right\rangle}=\frac{\hbar v_F}{2\sqrt{2}\Delta_0}. \tag{26}$$

Using the result of the BCS theory $\Delta_0=3.53k_BT_c/2$ [42,43], equation (26) is rewritten as

$$\bar{\rho} = 0.200315 \frac{\hbar v_F}{k_B T_c}, \tag{27}$$

where $T_c$ is the critical temperature of the SS.

As is well known, Pippard's coherence length is given by $\lambda_P = 0.1803 \hbar v_F / (k_B T_c)$ in the BCS theory [40, 42]. Equation (27) is in a good agreement with $\lambda_P$. Pippard's coherence length has been considered to be the mean distance of two electrons which consist of the Cooper pair [40, 41, 43]. Therefore, the maximum geminal of the BCS-SS can be reasonably regarded as the Cooper pair itself from the viewpoint of the present scheme using the RDM2 of the BCS-SS. Thus, we can say that the SS is the state such that the maximum geminal of the RDM2 of the SS is condensed as the Cooper pair in a number of $O(N)$.

Furthermore, it can be also confirmed that the BCS-SS actually meets the condition of the superconductivity defined in the previous paper [39] because the maximum geminal belongs to type (b) which is localized in a smaller region than the whole system in relative coordinates [39].

## 5. Discussions and concluding remarks

In this paper we deal with a long-standing suspicion as to whether the abnormal density is appropriate as the order parameter of SS or not. We prove that the cluster decomposition principle holds in the BCS-SS. Specifically, equation (19) can be proved true in the BCS-SS, so that the long-standing suspicion is dispelled at the level of BCS-SS. Although we discuss only the BCS state that is a special case but the most typical case of the superconductivity, this result means that the abnormal density may be an OP of the SS [39]. In other words, the present result may give a theoretical foundation of the facts that the abnormal density has been commonly used as the OP of the SS in the first-principles theory [5-36], and that the calculation results can explain the experiments, especially the critical temperature, reasonably well [5-36]. Superficially regarded, it seems to be strange to adopt the abnormal density as the OP of SS, because the abnormal density takes nonzero value only if the number of electrons in the system of interest is not invariant. However, it is not strange because the system of interest is an electron system that consists of superconducting electrons and may change the number of superconducting electrons.

In addition to the above, the cluster decomposition principle can be proven true from the other viewpoint. As is well known, the fluctuation of the particle number is given by

$$\langle \Theta_{\mathrm{BCS}} | \hat{N}^2 | \Theta_{\mathrm{BCS}} \rangle - \langle \Theta_{\mathrm{BCS}} | \hat{N} | \Theta_{\mathrm{BCS}} \rangle^2 = 4 n_{\max}^{(2)}, \tag{28}$$

in the BCS theory [42, 43]. Here we use the present result given by equation (15). On the other hand, in the previous paper [39], we have discussed the fluctuation of particle number in a general state not limited to the BCS state. Within the Hartree-Fock approximation it is calculated as [39]

$$\langle \Phi | \hat{N}^2 | \Phi \rangle - \langle \Phi | \hat{N} | \Phi \rangle^2 = \sum_{\zeta_1} \sum_{\zeta_2} \int \left\langle \psi^\dagger(\boldsymbol{r}_1\zeta_1)\psi^\dagger(\boldsymbol{r}_2\zeta_2) \right\rangle \langle \psi(\boldsymbol{r}_2\zeta_2)\psi(\boldsymbol{r}_1\zeta_1) \rangle d^3r_1 d^3r_2 \ . \tag{29}$$

If the cluster decomposition principle holds in the SS $|\Phi\rangle$, then equation (20) can be used. If that is the case, equation (29) is rewritten as

$$\langle \Phi | \hat{N}^2 | \Phi \rangle - \langle \Phi | \hat{N} | \Phi \rangle^2 = 2n_{\max}^{(2)} \ . \tag{30}$$

Difference between factors 4 and 2, which are seen in equations (28) and (30), seems to be caused by the Hartree-Fock approximation adopted in the previous paper [39]. This difference is not essential to the estimation of the fluctuation of particle number. The crucial point is that the fluctuation of the particle number is $O\left(n_{\max}^{(2)}\right)$ if the system is in the SS. A scale agreement between equation (28) and equation (30) also means that the cluster decomposition principle holds in the BCS-SS.

In the present paper, we also prove that the maximum geminal of the RDM2 of the BCS-SS can be regarded as the Cooper pair itself. Specifically, the maximum geminal is the spin singlet and spatially extends like s-wave, and the mean distance of electrons of the maximum geminal is in a good accordance with Pippard's coherence length of the BCS theory. It can be said that the SS is a many-electron state where thus-defined Cooper pair is condensed in a number of $O(N)$. In what follows, we shall reiterate the importance of this achievement.

Generally, there exist two kinds of pictures for the SS. The first one is to see the SS as the set of the Cooper pairs. In this picture, the superconductor is considered to be a set of the Cooper pairs which are condensed to the same energy level in a number of $O(N)$. Then, the Cooper pair in the case of the BCS-SS is spin singlet and extends like s-wave within the range of about Pippard's coherence length. The second picture is to describe the SS by means of the many-body wave function such as the BCS wave function. In this picture, the condensation of electron pairs to the same energy level is not explicitly expressed, but occupations of electron states within the free-electron Fermi surface are simply expressed. Unfortunately, the correspondence between these two pictures is not always trivial. This is because it is generally difficult in the many-electron system to define not only a single state (one-electron state) but also a pair state (two-electron state) due to the interaction between electrons. That is to say, it is generally difficult to construct one-electron picture (two-electron picture) in which many-electron state is described by a single configuration of a set of one-electron states (two-electron states). Therefore, it is difficult to define the Cooper pair from the SS that is of course a kind of many-electron state. In this paper, the problem is overcome by investigating the properties of the maximum geminal of the RDM2 of the SS.

Thus, the present work gives an answer to the issues “what is the OP of SS?”, and “what is the Cooper pair in the SS?” in the BCS-SS. Although the state to handle in this paper is the BCS-SS, this state is the most typical SS and therefore seems to be meaningful as a prototype of the various types of SSs.

**Acknowledgements**

This work was supported by JSPS KAKENHI grant numbers JP26400354, JP26400397, and JP16H00916.

**Appendix A:** Derivations of equations (8) and (9)

As is well known, the BCS wave function is given by [42]

$$|\Theta_{\mathrm{BCS}}\rangle = \prod_{\boldsymbol{k}} \left( u_{\boldsymbol{k}} + v_{\boldsymbol{k}} B_{\boldsymbol{k}}^{\dagger} \right) |0\rangle, \tag{A-1}$$

where $v_{\boldsymbol{k}}$ ($u_{\boldsymbol{k}}$) represents the probability amplitude such that $|v_{\boldsymbol{k}}|^2$ ($|u_{\boldsymbol{k}}|^2$) gives the probability of two electron state $|\boldsymbol{k}\uparrow, -\boldsymbol{k}\downarrow\rangle$ being occupied (unoccupied) in the BCS-SS, and are written as

$$|v_{\boldsymbol{k}}|^2 = \langle \Theta_{\mathrm{BCS}} | C_{\boldsymbol{k}\sigma}^{\dagger} C_{\boldsymbol{k}\sigma} | \Theta_{\mathrm{BCS}} \rangle, \tag{A-2}$$

and

$$|u_{\boldsymbol{k}}|^2 = 1 - |v_{\boldsymbol{k}}|^2, \tag{A-3}$$

respectively. Due to the normalization condition of $|\Theta_{\mathrm{BCS}}\rangle$, $u_{\boldsymbol{k}}$ and $v_{\boldsymbol{k}}$ are rewritten using a complex number $\alpha_{\boldsymbol{k}}$. We have

$$u_{\boldsymbol{k}} = \frac{1}{\left(1+|\alpha_{\boldsymbol{k}}|^2\right)^{\frac{1}{2}}}, \qquad v_{\boldsymbol{k}} = \frac{\alpha_{\boldsymbol{k}}}{\left(1+|\alpha_{\boldsymbol{k}}|^2\right)^{\frac{1}{2}}}. \tag{A-4}$$

Using the completeness of equation (3), the state $\left|\boldsymbol{r}\zeta, \boldsymbol{r}'\zeta'\right\rangle$ is written as

$$\left|\boldsymbol{r}\zeta, \boldsymbol{r}'\zeta'\right\rangle = \sum_{\boldsymbol{k}} \left|\boldsymbol{k}\uparrow, -\boldsymbol{k}\downarrow\right\rangle \left\langle \boldsymbol{k}\uparrow, -\boldsymbol{k}\downarrow \middle| \boldsymbol{r}\zeta, \boldsymbol{r}'\zeta'\right\rangle. \tag{A-5}$$

Here note that the completeness of two particle state is generally given by

$$\sum_{\boldsymbol{k}}\sum_{\sigma}\sum_{\boldsymbol{k}'}\sum_{\sigma'} \left|\boldsymbol{k}\sigma, \boldsymbol{k}'\sigma'\right\rangle \left\langle \boldsymbol{k}\sigma, \boldsymbol{k}'\sigma'\right| = 2. \tag{A-6}$$

However, due to the fact that the spin is fixed, the completeness of equation (3) is given by

$$\sum_{\boldsymbol{k}} \left|\boldsymbol{k}\uparrow, -\boldsymbol{k}\downarrow\right\rangle \left\langle \boldsymbol{k}\uparrow, -\boldsymbol{k}\downarrow\right| = 1. \tag{A-7}$$

Equation (A-5) gives the relation of the operators such that

$$\psi^\dagger(\boldsymbol{r}\zeta)\psi^\dagger(\boldsymbol{r}'\zeta') = \sum_{\boldsymbol{k}} \left\langle \boldsymbol{k}\uparrow, -\boldsymbol{k}\downarrow \middle| \boldsymbol{r}\zeta, \boldsymbol{r}'\zeta'\right\rangle B_{\boldsymbol{k}}^\dagger. \tag{A-8}$$

From equation (5), the RDM2 with respect to the BCS-SS is given by

$$\rho^{(2)}_{\Theta_{\rm BCS}}(\boldsymbol{r}_1\zeta_1\boldsymbol{r}_2\zeta_2, \boldsymbol{r}'_1\zeta'_1\boldsymbol{r}'_2\zeta'_2) = \frac{1}{2}\left\langle \Theta_{\rm BCS}\right| \psi^\dagger(\boldsymbol{r}'_1\zeta'_1)\psi^\dagger(\boldsymbol{r}'_2\zeta'_2)\psi(\boldsymbol{r}_2\zeta_2)\psi(\boldsymbol{r}_1\zeta_1)\left|\Theta_{\rm BCS}\right\rangle \tag{A-9}$$

Substituting Eqs (A-1) and (A-8) into equation (A-9), we have

$$\begin{aligned}\rho^{(2)}_{\Theta_{\rm BCS}}(\boldsymbol{r}_1\zeta_1\boldsymbol{r}_2\zeta_2, \boldsymbol{r}'_1\zeta'_1\boldsymbol{r}'_2\zeta'_2) = \frac{1}{2}\sum_{\boldsymbol{k}}\sum_{\boldsymbol{k}'}\Big\{ &\left\langle \boldsymbol{r}'_1\zeta'_1\boldsymbol{r}'_2\zeta' \middle| \boldsymbol{k}'\uparrow, -\boldsymbol{k}'\downarrow\right\rangle^* \left\langle \boldsymbol{r}_1\zeta_1\boldsymbol{r}_2\zeta_2 \middle| \boldsymbol{k}\uparrow, -\boldsymbol{k}\downarrow\right\rangle \\ &\times v_{\boldsymbol{k}'}^{*} v_{\boldsymbol{k}} \left\langle \tilde{\Theta}^{\boldsymbol{k}'}_{\rm BCS} \middle| \tilde{\Theta}^{\boldsymbol{k}}_{\rm BCS}\right\rangle \Big\},\end{aligned} \tag{A-10}$$

where the state $\left|\tilde{\Theta}^{\boldsymbol{k}}_{\mathrm{BCS}}\right\rangle$ is the state in which $\left|\boldsymbol{k}\uparrow,-\boldsymbol{k}\downarrow\right\rangle$ is excluded from $\left|\Theta_{\mathrm{BCS}}\right\rangle$. The explicit form is given by

$$\left|\tilde{\Theta}^{\boldsymbol{k}}_{\mathrm{BCS}}\right\rangle = \prod_{\boldsymbol{k}'(\neq\boldsymbol{k})}\left(u_{\boldsymbol{k}'} + v_{\boldsymbol{k}'}B^{\dagger}_{\boldsymbol{k}'}\right)\left|0\right\rangle. \qquad \text{(A-11)}$$

The inner product of $\left|\tilde{\Theta}^{\boldsymbol{k}}_{\mathrm{BCS}}\right\rangle$ and $\left|\tilde{\Theta}^{\boldsymbol{k}'}_{\mathrm{BCS}}\right\rangle$, which appears in equation (A-10), is calculated as

$$\left\langle\tilde{\Theta}^{\boldsymbol{k}'}_{\mathrm{BCS}}\middle|\tilde{\Theta}^{\boldsymbol{k}}_{\mathrm{BCS}}\right\rangle = u_{\boldsymbol{k}}^{*}u_{\boldsymbol{k}'} + v_{\boldsymbol{k}}^{*}v_{\boldsymbol{k}'}\delta_{\boldsymbol{k}\boldsymbol{k}'}. \qquad \text{(A-12)}$$

**Appendix B:** Proof for equation (13)

In this appendix, we shall show that the magnitude of $\sum_{\boldsymbol{k}'}\left|v_{\boldsymbol{k}'}\right|^2 u_{\boldsymbol{k}'}{}^2$ is $O(N)$. Using the gap parameter of the BCS theory, $\sum_{\boldsymbol{k}'}\left|v_{\boldsymbol{k}'}\right|^2 u_{\boldsymbol{k}'}{}^2$ can be rewritten as [42,43]

$$\sum_{\boldsymbol{k}'}\left|v_{\boldsymbol{k}'}\right|^2 u_{\boldsymbol{k}'}{}^2 = \frac{1}{4}\sum_{\boldsymbol{k}'}\frac{\Delta_{\boldsymbol{k}'}{}^2}{\xi_{\boldsymbol{k}'}{}^2 + \Delta_{\boldsymbol{k}'}{}^2}, \qquad \text{(B-1)}$$

where $\Delta_{\boldsymbol{k}'}$ is the gap parameter of the BCS theory and where $\xi_{\boldsymbol{k}'}$ is the free-electron energy shifted by the chemical potential $\mu$, i.e., $\xi_{\boldsymbol{k}'} = \varepsilon_{\boldsymbol{k}'} - \mu$. If we consider the case of s-wave superconductor, the gap parameter is approximated as $\Delta_0$ (=constant), and equation (B-1) becomes

$$\begin{aligned}\sum_{\boldsymbol{k}'}\left|v_{\boldsymbol{k}'}\right|^2 u_{\boldsymbol{k}'}{}^2 &= \frac{\Delta_0{}^2}{4}\sum_{\boldsymbol{k}'}\frac{1}{\xi_{\boldsymbol{k}'}{}^2 + \Delta_0{}^2} \\ &= \frac{\Delta_0{}^2}{4}\int_0^{\infty} d\varepsilon N(\varepsilon)\frac{1}{(\varepsilon-\mu)^2 + \Delta_0{}^2},\end{aligned} \qquad \text{(B-2)}$$

where $N(\varepsilon)$ is the density of states (DOS). If the weak coupling superconductor is considered, then the relation $\Delta_0 \ll \hbar\omega_D$ holds, where $\omega_D$ is the Debye frequency. Furthermore, the Debye temperature is generally much smaller than the Fermi temperature, i.e., $k_B T_D \ll k_B T_F$ that corresponds to the relation $\hbar\omega_D \ll \varepsilon_F \approx \mu$. Therefore, we have

$$\Delta_0 \ll \mu. \tag{B-3}$$

Taking this relation into account, the DOS in the integrant of equation (B-2) can be approximated as the constant value at the chemical potential. At that time, the integral of equation (B-2) can be done analytically, so that we get

$$\sum_{\boldsymbol{k}'} |v_{\boldsymbol{k}'}|^2 u_{\boldsymbol{k}'}{}^2 = \frac{\pi \Delta_0 N(\mu)}{4}. \tag{B-4}$$

Of course, the DOS is proportional to the volume of the system, which means that equation (B-4) is proportional to the particle number of the system.

**Appendix C:** Derivation of equation (22)

In this appendix, the derivation of equation (22) is explained. We consider the case where the attractive interaction between electrons is isotropic (BCS approximation [43]). The gap parameter is given by a constant one $\Delta_0$ from its definition. Using this gap parameter of the BCS theory, $u_{\boldsymbol{k}} v_{\boldsymbol{k}}$ is written as [42,43]

$$u_{\boldsymbol{k}} v_{\boldsymbol{k}} = \frac{1}{2} \left\{ \frac{\Delta_0{}^2}{\xi_{\boldsymbol{k}}{}^2 + \Delta_0{}^2} \right\}^{\frac{1}{2}}. \tag{C-1}$$

Then equation (21) is rewritten as

$$v^{\max}_{\Theta_{\mathrm{BCS}}} (\boldsymbol{r}\zeta, \boldsymbol{r}'\zeta') = \Xi_S (\zeta, \zeta') \sum_{\boldsymbol{k}} a_{\boldsymbol{k}} e^{i\boldsymbol{k}\cdot\boldsymbol{\rho}}, \tag{C-2}$$

where $a_{\boldsymbol{k}}$ is defined as

$$a_{\boldsymbol{k}} = \frac{1}{\sqrt{n^{(2)}_{\max}}} \frac{1}{2\Omega} \left( \frac{\Delta_0{}^2}{\xi_{\boldsymbol{k}}{}^2 + \Delta_0{}^2} \right)^{\frac{1}{2}}. \tag{C-3}$$

Here note that a phase factor $e^{i\varphi}$ which is attached to the BCS wave function [43] is omitted.

In general, the plane wave can be expanded in terms of the spherical harmonics [46]

$$e^{i\boldsymbol{k}\cdot\boldsymbol{\rho}} = 4\pi\sum_{l=0}^{\infty}\sum_{m=-l}^{+l} i^l j_l\left(k|\boldsymbol{\rho}|\right)Y_{lm}\left(\theta_{\boldsymbol{k}},\phi_{\boldsymbol{k}}\right)^* Y_{lm}\left(\theta_{\boldsymbol{\rho}},\phi_{\boldsymbol{\rho}}\right), \tag{C-4}$$

where $j_l\left(x\right)$ and $Y_{lm}\left(\theta_{\boldsymbol{k}},\phi_{\boldsymbol{k}}\right)$ are spherical Bessel function and spherical harmonics, respectively, and where $\theta_{\boldsymbol{k}}$ and $\phi_{\boldsymbol{k}}$ denote angles of $\boldsymbol{k}$ in the polar coordinates. Substituting equations (C-3) and (C-4) into equation (C-2), we can obtain equation (22). In the derivation, sum of $\boldsymbol{k}$ is converted into the integration with respect to $\boldsymbol{k}$ in a usual way of bulk system, and the orthogonality of the spherical harmonics is used.

**Appendix D:** Derivation of equation (25)

We shall calculate the mean distance between two electrons which consist of the maximum geminal $\left|\nu_{\Theta_{\mathrm{BCS}}}^{\max}\right\rangle$. This is given by

$$\left\langle|\boldsymbol{\rho}|^2\right\rangle = \frac{\sum_{\zeta_1}\sum_{\zeta_2}\int \nu_{\Theta_{\mathrm{BCS}}}^{\max}\left(\boldsymbol{\rho},\zeta_1,\zeta_2\right)^* |\boldsymbol{\rho}|^2 \nu_{\Theta_{\mathrm{BCS}}}^{\max}\left(\boldsymbol{\rho},\zeta_1,\zeta_2\right) d^3\rho}{\sum_{\zeta_1}\sum_{\zeta_2}\int \left|\nu_{\Theta_{\mathrm{BCS}}}^{\max}\left(\boldsymbol{\rho},\zeta_1,\zeta_2\right)\right|^2 d^3\rho}. \tag{D-1}$$

Substituting equation (C-2) into equation (D-1), equation (D-1) is written as

$$\left\langle|\boldsymbol{\rho}|^2\right\rangle = \frac{\sum_{\boldsymbol{k}}\left|\nabla_{\boldsymbol{k}} a_{\boldsymbol{k}}\right|^2}{\sum_{\boldsymbol{k}}\left|a_{\boldsymbol{k}}\right|^2}, \tag{D-2}$$

where following relations are used:

$$\int e^{i(\boldsymbol{k}-\boldsymbol{k}')\cdot\boldsymbol{\rho}} d^3\rho = \Omega\delta_{\boldsymbol{k},\boldsymbol{k}'}, \tag{D-3}$$

$$\sum_{\boldsymbol{k}} e^{i\boldsymbol{k}\cdot(\boldsymbol{\rho}-\boldsymbol{\rho}')} = \Omega\delta(\boldsymbol{\rho}-\boldsymbol{\rho}'). \tag{D-4}$$

From equation (C-3), $a_{\boldsymbol{k}}$ can be regarded as a function of $\xi\ (=\hbar^2k^2/(2m)-\mu)$, which is denoted as $a(\xi)$. Using the density of states $N(\varepsilon)$, the sum of $\boldsymbol{k}$ is converted by the integral with respect to $\xi$. Equation (D-2) is rewritten as

$$\left\langle |\boldsymbol{\rho}|^2 \right\rangle = \frac{\int_{-\mu}^{\infty} N(\xi+\mu) \left|\frac{\partial \xi}{\partial k}\right|^2 \left|\frac{\partial a(\xi)}{\partial \xi}\right|^2 d\xi}{\int_{-\mu}^{\infty} N(\xi+\mu) |a(\xi)|^2 \, d\xi}. \tag{D-5}$$

We consider the case where the attractive interaction between electrons is isotropic. Then, the gap parameter is given by a constant one $\Delta_0$ from its definition. The function $|a(\xi)|^2$ included in the integrand of the denominator is written as

$$|a(\xi)|^2 = \frac{1}{n_{\max}^{(2)}} \frac{1}{4\Omega^2} \left( \frac{\Delta_0^{\ 2}}{\xi^2 + \Delta_0^{\ 2}} \right). \tag{D-6}$$

This function has a sharp peak about the origin $\xi = 0$ the full width half maximum (FWHM) of which is $2\Delta_0$. Since $\Delta_0$ is much smaller than the chemical potential $\mu$, the DOS included in the integrand may be approximated as a constant value $N(\mu)$. Thus, the denominater of equation (D-5) is given by

$$\begin{aligned} \int_{-\mu}^{\infty} N(\xi+\mu) |a(\xi)|^2 \, d\xi &\approx \frac{N(\mu)}{4\Omega^2 n_{\max}^{(2)}} \int_{-\mu}^{\infty} \frac{\Delta_0^{\ 2}}{\xi^2 + \Delta_0^{\ 2}} d\xi \\ &= \frac{N(\mu)\Delta_0}{4\Omega^2 n_{\max}^{(2)}} \left\{ \frac{\pi}{2} - \arctan\left( \frac{-\mu}{\Delta_0} \right) \right\}. \end{aligned} \tag{D-7}$$

Since $\Delta_0 \ll \mu$, $\arctan(-\mu/\Delta_0)$ is reasonably approximated as $-\pi/2$. Then equation (D-7) is given by

$$\int_{-\mu}^{\infty} N(\xi+\mu) |a(\xi)|^2 \, d\xi \approx \frac{N(\mu)\Delta_0 \pi}{4\Omega^2 n_{\max}^{(2)}}. \tag{D-8}$$

In a similar way to the denominator, the function $|\partial a(\xi)/\partial \xi|^2$ included in the numerator of equation (D-5) takes nonzero value only around the origin $\xi = 0$. Namely, it is symmetrical about $\xi = 0$, and has two peaks the distance of which are $\sqrt{2}\Delta_0$. Since $\Delta_0 \ll \mu$, the numerator of equation (D-5) is reasonably approximated as

$$\int_{-\mu}^{\infty} N(\xi+\mu)\left|\frac{\partial\xi}{\partial k}\right|^2 \left|\frac{\partial a(\xi)}{\partial\xi}\right|^2 d\xi \approx N(\mu)\left|\left(\frac{\partial\xi}{\partial k}\right)_{\xi=0}\right|^2 \int_{-\mu}^{\infty}\left|\frac{\partial a(\xi)}{\partial\xi}\right|^2 d\xi$$
$$\approx \frac{N(\mu)(\hbar v_F)^2}{4\Omega^2 n_{\max}^{(2)}}\frac{\pi}{8\Delta_0},$$

(D-9)

where the approximation such that $\arctan(-\mu/\Delta_0)\approx -\pi/2$ , and the relation $(\mu/\Delta_0)^2 \gg \mu/\Delta_0 \gg 1$ are used.

Thus, substituting the above results into equation (D-5), we finally get

$$\left\langle |\boldsymbol{\rho}|^2 \right\rangle = \frac{\hbar^2 {v_F}^2}{8{\Delta_0}^2}. \qquad \text{(D-10)}$$